\documentclass[prb,preprint]{revtex4}
\usepackage{graphicx}
\makeatletter
\usepackage{epsfig}
\def\@dotsep{4.5}
\begin{document}
\title{Calculations of Trapping and Desorption in Heavy Atom Collisions with Surfaces}

\author{Guoqing Fan and  J. R. Manson}
 \affiliation{Department of Physics and Astronomy,Clemson University, Clemson,
SC, 29634}
\email{jmanson@clemson.edu}   
\date{\today}

\begin{abstract}
Calculations are carried out for the scattering of heavy  rare gas atoms with surfaces using a recently
developed classical theory that can track particles trapped in the physisorption potential well and follow them
until ultimate desorption.  Comparisons are made with recent experimental data for xenon scattering from molten
gallium and indium, systems for which the rare gas is heavier than the surface atoms.  The good agreement with
the data obtained for both time-of-flight energy-resolved spectra and for total scattered angular distributions
yields an estimate of the physisorption well depths for the two systems.
\end{abstract}


\maketitle

\section{Introduction}

The scattering of thermal and hyperthermal energy heavy rare gas particles  from surfaces composed of atoms of
smaller mass presents an interesting problem because in these cases most of the incident beam is trapped at the
surface and then subsequently desorbed after spending a substantial amount of time in the physisorption
potential well.  This large initial trapping fraction arises for two reasons,  the mass ratio is large favoring
forward scattering events in the initial collisions with surface atoms and for such systems the physisorption
well depths are typically large.  A simple and useful assumption that is often applied in such cases, which was
first suggested by Maxwell~\cite{Maxwell}, is to assume that the initially trapped fraction eventually desorbs
in equilibrium with the temperature of the surface.  However, such an assumption is only approximate and
experiments have shown that in many cases there can be substantial deviations from equilibrium in the
trapping-desorption fraction.~\cite{Nathanson,Sibener-JCP-03,Morris-JCP-2003}

The purpose of this paper is to carry out calculations with a recently developed theory of gas-surface
scattering  that can not only treat direct scattering but can also track the initially trapped particles and
follow them until they are ultimately desorbed.~\cite{Fan-07} The gas particles trapped in a physisorption
potential can have either positive or negative total energies.  Those with negative energy particles are truly
bound to the surface and will not desorb until they receive enough energy from the surface in later collisions
to enable their escape back into the continuum.  The positive energy particles are those that remain moving
parallel to the surface with sufficiently large velocities that the total energy is positive, but they are
directed at such grazing angles inside the well that they cannot escape.  The positive energy part is sometimes
called the chattering fraction.

The theory applied here treats the trapped particles, those with both positive and negative energy, as a series
of successive collisions with the surface while bound in the well.  Energy transfer occurs only with the
repulsive wall, while the attractive part of the potential is considered rigid and gives rise to specular
reflection of those particles which have neither sufficient energy nor sufficiently large reflection angles to
escape.  In fact, the attractive $C_3/z^3$ potential of the asymptotic Van der Waals potential is rigid because
it arises from a sum of pairwise atomic Van der Waals potentials over all atoms in the bulk half-space and hence
all vibrational motions average to zero.  The fact that the scattering is treated as successive collisions
allows for an iterative treatment that can be extended to very large trapping times before essentially all of
the initially trapped particles are scattered back into the continuum.

The iterative approach to trapping-desorption used here is largely independent of the scattering potential,  and
for the heavy gas atoms considered we use the scattering model initially treated by Brako and Newns  for a
smooth surface with vibrational displacement corrugations due to the thermal motions of the underlying substrate
atoms.~\cite{model1,model2,model4}  This model has been extended and shown to explain atomic and molecular
surface scattering experiments in which the direct scattering contributions are
dominant.~\cite{Muis-97,Hayes3}
The work in this paper shows that this same scattering potential model can explain experiments in which there is
little or no direct scattering and where the trapping-desorption fraction is dominant.

The calculations are compared with experimental data for the scattering of well-defined incident beams of Xe
atoms  colliding with molten Ga and In surfaces.~\cite{Nathanson}  Both total angular distributions and
energy-resolved time-of-flight spectra are considered.  Comparisons with the angular distributions from a Ga
surface confirms earlier observations that collective effects require an effective mass that is somewhat larger
than the atomic mass for the liquid Ga.  Good agreement is obtained with both angular distributions and
energy-resolved spectra which enables an estimation of the physisorption potential well for both the Xe/Ga and
the Xe/In systems.

\section{THEORY}\label{theory}

In a surface scattering event the basic collision is often described in terms of a differential  reflection
coefficient ${dR({\bf p}_f, {\bf p}_i; T_s)}/{{dE_f}{d\Omega _f}}$ which gives the probability that a particle
with well-defined incident energy impinging on the surface at a given angle will be scattered into the small
interval $dE_f$ centered about the final value $E_f$ and the small interval $d\Omega _f$ centered about the
final solid angle $\Omega _f$.
The collision with the repulsive surface potential occurs inside the physisorption well which is assumed to  be
a square well of uniform depth $D$ and width $b$, the width being unimportant as long as it is larger than the
vibrational displacements of the selvage region of the surface corrugation.  This means that the energy
$E_q^\prime$ of a particle inside the well compared to the energy $ E_q$ of an equivalent particle outside the
well  is given by
\begin{eqnarray}  \label{M1}
E_q^\prime ~=~ E_q ~+~ D
~,
\end{eqnarray}
where the label $q$ can denote an incident,  final or intermediate state.  All of the additional energy from
the well depth goes only into increasing the normal momentum component $p_{qz} $ of the particle
\begin{equation} \label{M2}
p'^2_{qz} = p^2_{qz} + 2mD
~.
\end{equation}
Eqs.~(\ref{M1}) and~(\ref{M2}) describe a refraction of the incident particles towards more normal angles
inside the well.   The connection between the differential reflection coefficient calculated inside the well and
that describing the fraction which escape into the continuum desorption states outside the well is given by a
Jacobian that is calculated from Eqs.~(\ref{M1}) and~(\ref{M2}).

The collision process proceeds as follows: the incident beam of particles approaches the surface, enters  the
physisorption well and proceeds to strike the repulsive wall where the particles exchange energy and momentum.
Upon reflection from the repulsive wall some particles lose sufficient energy that they are trapped in the
negative energy states, some are reflected at angles and energies that cause them to enter the positive energy
chattering states, while the remainder escape the surface as the direct scattering fraction. The trapped
particles are specularly reflected at the attractive wall of the well and proceed to have a second collision
with the repulsive potential.

The whole process repeats, but after each collision a fraction of the trapped particles are elevated into
angular  and energy ranges that allow them to escape and desorb into the continuum states.

Thus the trapped particles act as the source for subsequent collisions  in a process that can be characterized
by the following equation: 
\begin{eqnarray} \label{iter}
\frac{dR({\bf p}_f,{\bf p}_i)}{d  {E}_f d \Omega_f}  ~=~ \frac{dR^0({\bf p}_f,{\bf p}_i)}{d  {E}_f d \Omega_f}
~+~ \int d E_b d \Omega_b ~ \frac{dR^0({\bf p}_f,{\bf p}_b)}{d  {E}_f d \Omega_f} ~ \frac{dR^0({\bf p}_b,{\bf
p}_i)}{d  {E}_b d \Omega_b}
\\ \nonumber
~+~ \int d E_b d \Omega_b ~ \frac{dR^0({\bf p}_f,{\bf p}_b)}{d  {E}_f d \Omega_f} ~ \frac{dR^1({\bf p}_b,{\bf
p}_i)}{d  {E}_b d \Omega_b} ~+~ \ldots
\\ \nonumber
~+~ \int d E_b d \Omega_b ~ \frac{dR^0({\bf p}_f,{\bf p}_b)}{d  {E}_f d \Omega_f} ~ \frac{dR^{n-1}({\bf
p}_b,{\bf p}_i)}{d  {E}_b d \Omega_b}
~,
\end {eqnarray}
where $ {dR^n({\bf p}_b,{\bf p}_i, T_S )} / { d E_b \: d \Omega_b} $ is the differential reflection coefficient
 for the distribution of particles remaining trapped in the bound states after $n$ collisions and the
intermediate integrations in the higher order terms are carried out only over angles and energies that pertain
to the trapped fraction.
The process described in Eq.~(\ref{iter}) is readily developed into an iterative procedure that can be continued to very high orders until the fraction of particles remaining trapped is arbitrarily small.   The details, including how to handle the differences between the positive and negative energy trapped fractions and how to calculate trapping times, are presented elsewhere.~\cite{Fan-07}

This procedure depends on the choice of differential reflection coefficient and for  the calculations presented
here we choose the model first developed by Brako and Newns for a smooth surface whose vibrating surface
corrugations are a linear response to the underlying thermal vibrations of the underlying atoms. This is a
differential reflection coefficient whose specific form is~\cite{model1,model2,model4}
\begin{equation} \label{M5}
\frac{dR({\bf p}_f,{\bf p}_i; T_s)}{dE_f d\Omega _f} = \frac {m^2 {v_R^2 } \left| {\bf p}_f \right|} {4\pi ^3
\hbar ^2 p_{iz} {S_{u.c.} } N_D^0 } \left| {\tau _{fi} } \right|^2 \left( \frac{\pi }{k_B T_s \Delta E_0 }
\right)^{3/2} \exp \left\{ - \frac{(E_f - E_i  + \Delta E_0 )^2+{2v_R^2 {\bf P}^2} }{4k_B T_s \Delta E_0 }
\right\} ~,
\end {equation}
where $\Delta E_0  = ({\bf p}_f - {\bf p}_i )^2 /2M $ is the recoil energy, $p_{iz}$ is the $z$ component of the
incident momentum, $k_B$ is the Boltzman constant, $T_s$ is the temperature of the surface, $\left| {\tau _{fi}
} \right|$ is the form factor of the scattering, $N_D^0$ is the normalization coefficient, ${\bf P}  = {\bf P}_f
- {\bf P}_i $ is the  parallel momentum exchange and $S_{u.c.}$ is the area of a surface unit cell. $v_R$ is a
velocity of sound parallel to the surface whose value is estimated to be in the range of the Rayleigh sound
velocity.  It is a weighted average over all surface vibrational modes and can be calculated if the surface
phonon spectral density at the classical turning point is known,  however, in this paper $v_R$ is treated as a
parameter as is often the case~\cite{model2,model3}.

The form factor $\left| {\tau _{fi} } \right| ^2$ is the squared amplitude of the  transition matrix element for
the elastic interaction potential extended off the energy shell for inelastic interactions in which energy is
transferred during the collision.~\cite{Celli-Himes-Manson} For all calculations presented here we use the hard
repulsive wall limit derived from the class of Mott-Jackson-like potentials which is~\cite{Goodman-76}
\begin{equation} \label{tfi}
 \tau _{fi}= 4p_{fz}p_{iz}/m
\end {equation}
This expression for the form factor has proven to be satisfactory for calculations of atomic and molecular
scattering from surfaces in which the direct scattering component was
dominant.~\cite{Muis-97,Muis-99,Iftimia1,Iftimia2,Moroz1,Moroz2,Moroz3,Ambaye1,Ambaye2,Ambaye3,Hayes1,Hayes2,Hayes3}

The quantity to be compared with the time-of-flight spectra after conversion to energy  transfer is the
differential reflection coefficient.  There is also a correction for the $1/v_f$ velocity dependence of the
experimental detector and for the comparisons made here this correction was applied to the data.   The measured
angular distributions are summed over all final energies and are compared with the differential reflection
coefficient per unit final solid angle which is the integral over all final energies of the differential
reflection coefficient according to
\begin{equation} \label{M4}
\frac{dR}{d \Omega_f}  ~=~ \int_0^\infty d E_f ~ \frac{dR({\bf p}_f ,{\bf p}_i; T_s)}{dE_f d\Omega _f}
\end {equation}

Collision times can be calculated by using the differential reflection coefficient of the  trapped particles as
a distribution function to calculate average times per collision iteration and the details are given in
Ref.~[\cite{Fan-07}].  For a potential with a square well the collision times are proportional to the well width
$b$ and for all calculations presented here we have chosen $b=3$~\AA.  The  collision times are also calculated
under the assumption that effectively all of the initially trapped particles are desorbed when a fraction of
only 1\% of the initially incident particles remain in the well.

\section{Comparison with Data}\label{result}

\subsection{Xenon on Indium}  \label{xein}

\begin{figure}
\includegraphics[width=5.0in]{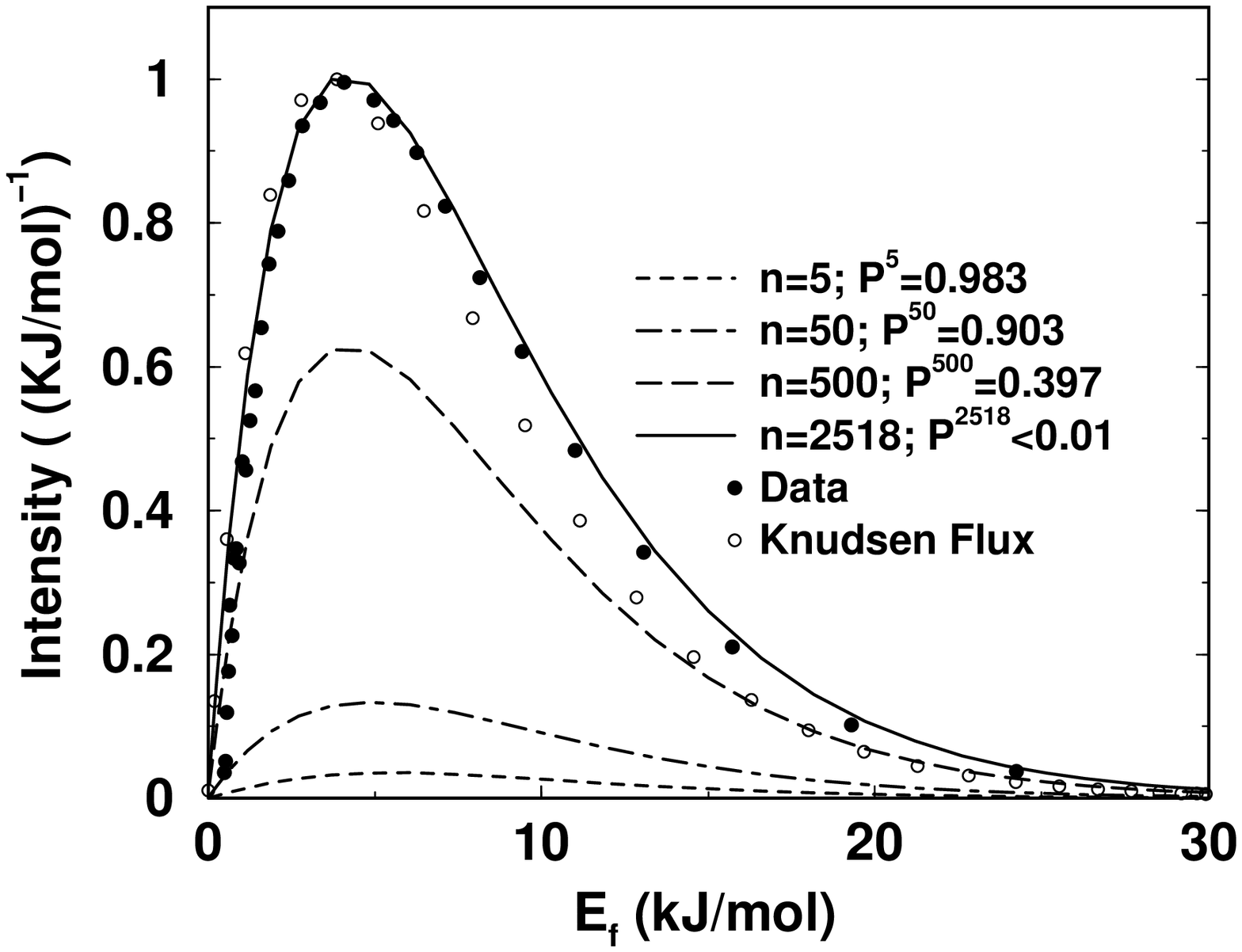}
\caption{ Energy-resolved scattering intensity for a  6 kJ/mol  beam of Xe atoms incident on a molten In surface
with an incident polar angle of 55$^\circ$ with respect to the normal. The final angle is 65$^\circ$  and the
surface temperature is 436 K. The data from Ref.~[\cite{Nathanson}] are shown as solid cirles, a Knudsen
equilibrium distribution is shown as open circles and the calculation with well depth 180 meV and $v_R=400$ m\/s
is shown as the solid curve.  } \label{energy-iter}
\end{figure}

Calculations for the energy-resolved intensity spectrum of a well-defined beam of Xe atoms scattering  from a
molten In surface are presented in Fig.~\ref{energy-iter}.  The incident energy is $E_i=6$ kJ/mol (62 meV) and
the incident angle is $\theta_i=55^\circ$.    The final polar angular position of the detector is
$\theta_f=65^\circ$ and lies in the scattering plane in the quadrant opposite to that of the incident beam. The
surface temperature is 436 K which is 6 K above the melting point.  The calculations are compared with recent
measurements by Nathanson et al.~\cite{Nathanson}   For reference, an equilibrium Knudsen flux distribution
given by
\begin{equation}  \label{MB}
\frac{dP^{K}({\bf p}_i, T_G )}{ d E_i \: d \Omega_i}
\; = \;
\frac{E_i \cos \theta_i}{\pi (k_B T_G)^2}  \exp \left\{   \frac{-E_i}{k_B T_G}  \right\}
  \; ,
\end{equation}
is also shown.

There is no evidence in the data for a distinct direct scattering peak, and this is  consistent with the
calculations which give an initial trapping fraction of approximately 99\% and virtually no direct scattering
for these incident conditions. The lack if significant direct scattering is not unexpected for two reasons:
first, the well depth is expected to be large compared to the incident energy favoring trapping processes, and
second, because the incident energy is not much larger than the most probable energy of the equilibrium
distribution at this surface temperature. It is interesting that, although the scattered spectrum is roughly of
the same form as the equilibrium Knudsen distribution, there are significant differences especially on the high
energy side where the scattered particles actually gain energy from the surface.

The calculations with $v_R=400$ m\/s
agree quite well with the data, significantly better than the agreement with the equilibrium
curve.
Calculations carried out for values of the parameter $v_R$ that are roughly 25\% larger or smaller are nearly  the same indicating that the results are not highly dependent on the choice of this
parameter.
The well depth predicted by the calculation is $D= 180$ meV.  The well depth for this system has not
been measured by independent experiments but this value is in the expected range for potentials of Xe
interacting with other metal surfaces.~\cite{Cole-Ihm-Bruch,Bruch}  It is interesting to note that an attempt to
subtract an equilibrium distribution from the experimental data, and then to match the residual intensity with
the direct scattering calculated from Eq.~(\ref{M5}) results in a significantly smaller estimate for the well
depth $D$.

Fig.~\ref{energy-iter} shows how the calculated spectrum develops as the trapped  particles
make continued interactions with the surface inside the well.  Initially, almost all of the incident particles
are trapped.  After 5 collision iterations, labeled $N=5$ in Fig.~\ref{energy-iter}, more than 98\% of the
incident particles still remain trapped.  Only after more than 2500 collisions are the particles essentially all
desorbed, at which point the dominant trapping-desorption spectrum explains very well the measured intensity.
The trapping time, assuming an arbitrary cut-off of less than 1\% of the incident particles remaining in the well and assuming a well width
$b=3$~\AA, is $\tau={ 2.2e^{-9}}$ s.

\subsection{Xenon on Gallium}\label{xega}

\begin{figure}
\includegraphics[width=5.0in]{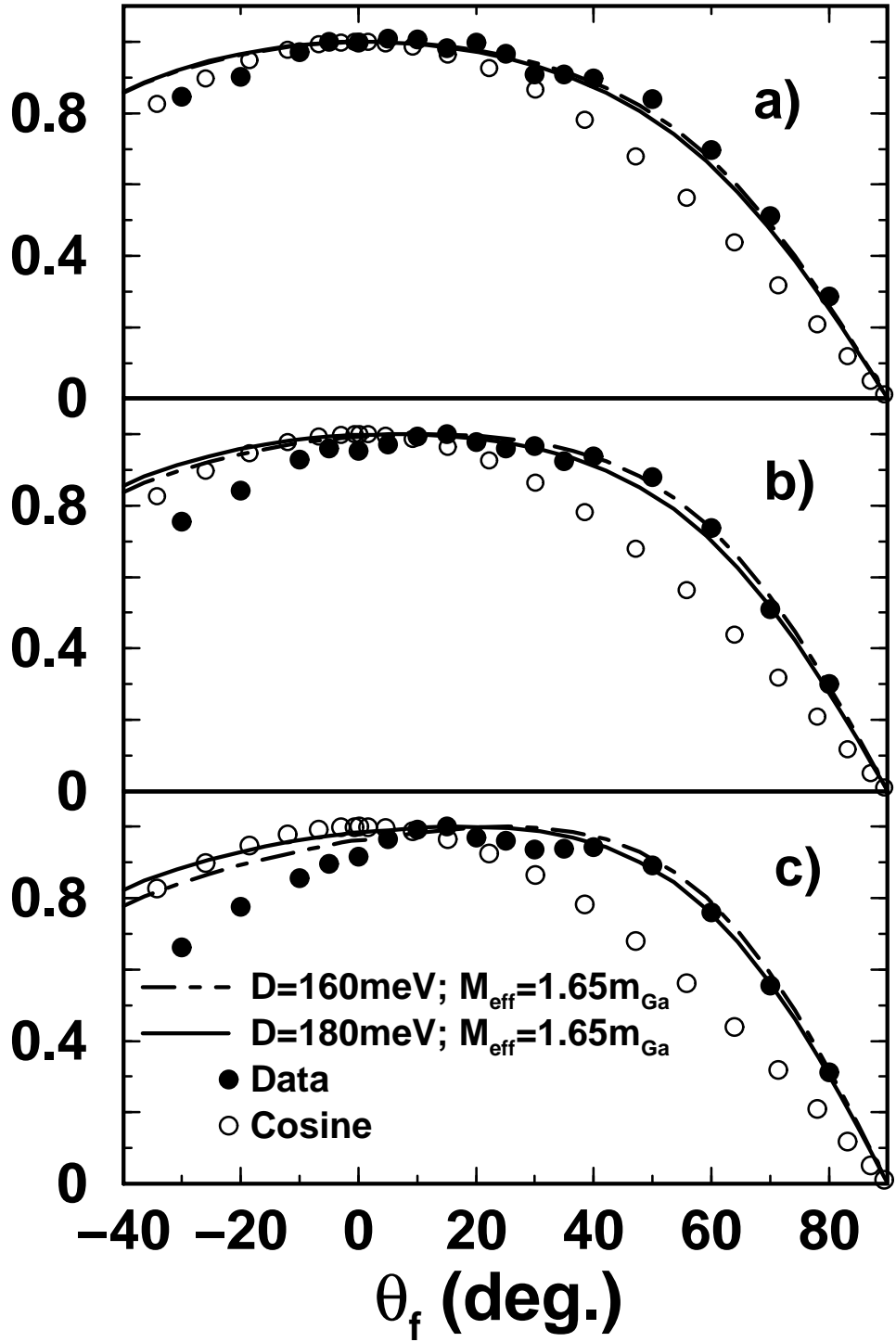}
\caption{ In-plane angular distributions for Xe scattering from  a molten Ga surface.  The incident energy is 6
kJ/mol, the incident angle is $55^\circ$ and the surface temperatures are: a) 308 K, b) 436 K and c) 586 K.  Two
calculations for well depths $D= 160$ and 180 meV are shown as long-dashed and solid curves, respectively.   The
data from Ref.~[\cite{Nathanson}] are shown as solid circles and an equilibrium $\cos \theta_f$ curve is shown
as open circles. } \label{xega308-vr400}
\end{figure}

Three examples of the  angular distributions scattered by a well-defined  beam of Xe incident on a  molten Ga
surface are shown in Fig.~\ref{xega308-vr400}.  The incident energy is $E_i=6$ kJ/mol, the incident  polar angle
with respect to the surface normal is $\theta_i=55^\circ$  and the three different temperatures are $T_S=308$,
436 and 586 K. The detector is always positioned in the scattering plane and final polar angles with values less
than zero indicate that the detector is in the same quadrant as the incident beam. The experimental data is
taken from Ref.~[\cite{Nathanson}] and an equilibrium $\cos \theta_f$ curve is also shown.

The calculations indicate that at all final angles trapping-desorption is the dominant process and direct
scattering is negligible.  At all three temperatures the calculated curves agree reasonably well with the data,
especially for positive $\theta_f$.  For final angles in the same quadrant as the incident beam the agreement is
somewhat less good and those discrepancies are accentuated as the temperature increases.  However, at all
temperatures the calculations agree substantially better with the measurements than does the equilibrium curve.

Calculations are shown for two different well depths, 160 and 180 meV.  Although there is not a large difference
between the two, the calculations for 160 meV appear to agree slightly better with the data.

All calculations shown in Fig.~\ref{xega308-vr400} were carried out using an effective mass for Ga that  is 1.65
times that of the Ga atomic mass.  The use of an effective mass is consistent with two other analysis of rare
gas scattering from liquid Ga.  An earlier analysis of Ne and Ar scattering from several liquid metals at higher
energies where direct scattering was the dominant contribution showed that agreement with the data for Ga could
not be obtained without using a larger effective mass, whereas for liquid In and Bi no increase of the effective
mass above the atomic mass was necessary.~\cite{Dai-EffectiveMass}  More recent and much more extensive data,
when analyzed with similar theoretical models, showed the same anomaly: a larger effective mass for Ga was
needed to explain scattering measurements of the rare gases Ne, Ar, Kr and Xe but not for liquid In and
Bi.~\cite{Hayes4} Although the physical origins of the need for a larger effective mass for Ga are not clearly
understood, this is consistent with independent measurements that show a large degree of residual layered
ordering in liquid Ga that persists even for temperatures higher than considered here whereas for the other two
low melting point metals such layering behavior is much less pronounced.~\cite{Pershan-LiquidMetals}

Just as for the case of the Xe/In interaction potential discussed above, the value of the physisorption well
depth is not known for Xe/Ga.  However, the prediction obtained here of $D \approx 160-180$ meV is in line with
values estimated for Xe-metal interaction potentials.~\cite{Cole-Ihm-Bruch,Bruch}  An earlier analysis of the present
data obtained by estimating the trapping-desorption fraction by subtracting  an equilibrium cosine distribution
from the data produced an estimate of the well depth of a rather unrealistically small value of approximately
100 meV.  This current calculation seems not only to explain the scattering data better, but produces a more
realistic value of the well depth.

\begin{figure}
\includegraphics[width=5.0in]{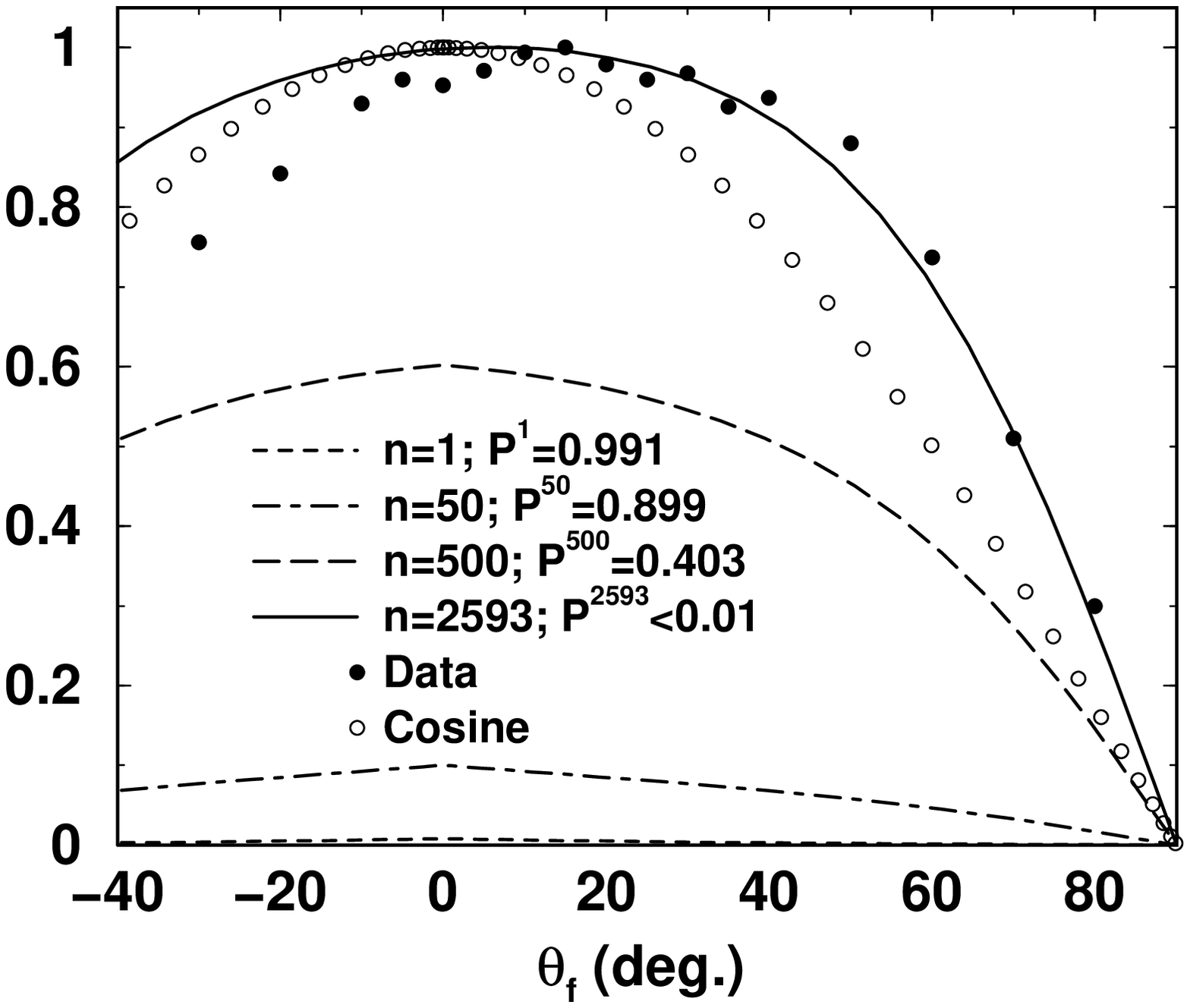}
\caption{ Same as in Fig.~\ref{xega308-vr400} b) for the calculation with $D=180$ meV but in addition giving
several calculated curves for collision iteration numbers showing partial desorption of the trapped fraction.  }
\label{xega308}
\end{figure}

Fig.~\ref{xega308} shows the calculation and data for $T_S=436$ of Fig.~\ref{xega308-vr400}
b) but also shows the development of the trapping-desorption intensity as  a function of collision iteration
number.  After the initial collision essentially all of the incident particles are trapped.  Even after 500
collision iterations there still remains nearly half the particles in the trapped states.  Only after more than
2500 iterations are essentially all particles desorbed, corresponding to a trapping time of $\tau={ 2.19
e^{-9}}$ s under the assumption of a well width of 3~\AA.

The lower panel of Fig.~\ref{xega308} shows the effect of the effective mass.  The surface temperature is 308 K
and two calculations similar to Fig.~\ref{xega308-vr400} a) are shown for two values of the parameter $v_R$, 300
and 400 m/s.  These two calculations are very similar, again showing the the results do not depend strongly on
this parameter.  Also shown, however, is a calculation with the Ga effective mass set equal to the atomic mass.
The considerable disagreement in comparison with the data indicates that this experiment is quite sensitive to
the effective mass even though the trapped atoms spend a considerable time residing in the bound states and
continually interacting with the surface potential.

\section{CONCLUSIONS}\label{conclusion}

In this paper we have applied a straightforward classical theoretical model for scattering from a smooth surface
to the case in which the trapping-desorption processes are the dominant scattering mechanism.  This same smooth
surface model has been shown in several previous studies to give good explanations of the scattering behavior
when direct scattering is the dominant mechanism in atomic and molecular surface
scattering.~\cite{Muis-97,Hayes3}

For the present work, we consider situations with large projectile-to-surface mass ratios and deep physisorption
potential well depths, conditions in which the initial collision with the surface results in nearly complete
trapping.  The subsequent desorption of the physisorbed fraction is modeled by an iterative process in which the
trapped particles continue to make collisions while inside the well and at each collision a small fraction
escape.  Eventually, the calculation can be followed to large iterations until all particles desorb.

Calculations are compared with recent data published for thermal energy beams of xenon scattering from molten
indium and gallium.  For both of these systems it was noted that the scattering was nearly entirely
trapping-desorption in nature, but both the measured energy-resolved and angular distribution spectra resembled
only approximately the distributions expected for a gas escaping from a surface in equilibrium.  However, the
calculations reported here explain the data quite well and give predictions of well depths for both systems in
the expected range of 160-180 meV.

In addition to the Xe/In and Xe/Ga systems presented here, numerous additional calculations have been carried
out with this theoretical model and compared with available data for the scattering of thermal energy rare gases
at liquid metal surfaces.~\cite{Fan-Thesis}  The calculations generally agree well with the data and produce
well depth predictions that are reasonable.  The calculational method also provides a means of examining the
behavior of the scattering process while the initially physisorbed particles are desorbing, and it gives
estimates of the trapping times.  Thus this theory seems to be a useful model for simulating atomic and
molecular scattering under conditions in which the trapping-desorption fraction is significant.

\vspace{1cm}
\noindent
{\bf Acknowledgements} \\

\noindent
This work was supported by the US Department of Energy under grant number
DMR-FG02-98ER45704.


\newpage
\begin{thebibliography} {99}


\bibitem{Maxwell} J. C. Maxwell, Philosophical Transactions of the Royal
Society of London, {\bfseries Vol. 170} 251 (1879).

\bibitem{Sibener-JCP-03} K. D. Gibson, N. Isa, and S. J. Sibener, J. Chem. Phys. {\bf 119} 13083 (2003).

\bibitem{Morris-JCP-2003}   B. Scott Day, Shelby F. Shuler, Adonis Ducre, and John R. Morris, J. Chem. Phys. {\bf 119} 8084
(2003).
\bibitem{Nathanson} W. R. Ronk, D. V. Kowalski, M. Manning and G. Nathanson
J. Chem. Phys. {\bf 104}, 4842 (1996).


\bibitem{Fan-07} Guoqing Fan and J. R. Manson, to be published; arXiv:0804.1776.

\bibitem{model1} R. Brako, D. M. Newns, Phys. Rev. Lett. {\bf 48}, 1859 (1982);

\bibitem{model2} R. Brako, D. M. Newns, Surf. Sci. {\bf 123}, 439 (1982) .

\bibitem{model4} J. R. Manson, Phys. Rev. B {\bf 43}, 6924 (1991).

\bibitem{Muis-97} A. Muis and J. R. Manson, J. Chem. Phys. {\bf 107} 1655 (1997).

\bibitem{Hayes3} W. W. Hayes,  Hailemariam Ambaye  and J. R. Manson, Journal of Physics:  Condens.  Matter {\bf 19}, 305007 (2007)


\bibitem{model3} H.-D. Meyer and R. D. Levine, Chem. Phys.  {\bf 85}, 189 (1984).

\bibitem{Celli-Himes-Manson} J. R. Manson, V. Celli and D. Himes , Phys. Rev. B {\bfseries 49}
2782 (1994) .

\bibitem{Goodman-76} F. O. Goodman and H. Y. Wachman, {\em Dynamics of Gas-Surface
Scattering} (Academic Press, New York, 1976).


\bibitem{Hayes4} W. W. Hayes and J. R. Manson, J. Chem. Phys. {\bf 127 }, 164714 (2007).


\bibitem{Dai-EffectiveMass} Jinze Dai and J. R. Manson, J. Chem. Phys.  {\bf 119}, 9842 (2003).


\bibitem{Pershan-LiquidMetals} M. Regan, E. Kawamoto, S. Lee, P. Pershan, N. Maskil, M. Deutsch,
O. Magnussen, B. Osko, L. Berman, Phys. Rev. Let. {\bf  75}, 2498 (1995).

\bibitem{Cole-Ihm-Bruch}  G. Vidali, G. Ihm, H.-Y. Kim and M. W. Cole, Surf. Sci. Rep. {\bf 12}, 133-181 (1991).

\bibitem{Bruch}  L. Bruch, M. W. Cole and E. Zaremba, "{\em Physical Adsorption: Forces and Phenomena}", (Oxford University Press, Oxford, 1997); Corrected reprinting by (Dover Publications, Mineola, New York, 2007).

\bibitem{Fan-Thesis} Guoqing Fan, Ph.D. dissertation, Clemson University (2007), unpublished.





\end{thebibliography}
\end{document}